**Visuality in a Cross-disciplinary Battleground: Analysis of Inscriptions in Digital Humanities Journal Publications**

Rongqian Ma
School of Computing and Information, University of Pittsburgh, PA 15213
Room 602, 135 N. Bellefield Ave., Pittsburgh, PA 15213
(732) 210-7127; rom77@pitt.edu

Kai Li*
School of Information Resource Management, Renmin University of China, Beijing, China
59 Zhongguancun St, Haidian District, Beijing, China, 100872
+86 13161285125; kai.li@ruc.edu.cn
*Corresponding author

**Abstract**

Like the old saying, "a graph is worth a thousand words," the non-verbal language, encapsulated in the concept of *inscription*, is a fundamental rhetorical device in the construction of knowledge represented by research outputs. As many inscriptions are deeply situated in a scientific and data-driven research paradigm, they can be used to understand the relationships between research traditions involved in the field of digital humanities (DH), a highly cross-disciplinary field that is frequently regarded as a battleground between these distinct research traditions, especially the humanities and STEM fields. This paper presents a quantitative, community-focused examination of how inscriptions are used in English-language research articles in DH journals. We randomly selected 252 articles published between 2011 and 2020 from a representative DH journal list, and manually coded and classified inscriptions and author domains in these publications. We found that inscriptions have been increasingly used during the past decade and their uses are more intensive in publications led by STEM authors comparing to other domains. This study offers a timely survey of the disciplinary landscape of DH from the perspective of inscriptions and sheds lights on how different research approaches collaborated and combated in the field of DH.

**Keywords**

digital humanities, visualization, inscriptions, quantitative science studies, disciplinarity

**Introduction**

The development of digital humanities (DH) has long been regarded as a "battleground" between various research domains and conventions, especially the humanities traditions and those from the STEM (i.e., science, technology, engineering, mathematics) fields. While cross-disciplinary dialogue is constantly embodied in the digital humanities (e.g., "Big Tent DH"; Svensson, 2010, 2012), tensions also exists. The *computational turn* has offered the humanities new research opportunities but also ignited competing attitudes towards digital methods and technologies among the communities (Berry, 2012). Proponents argued for the enabling effects of digital technologies and methods on transforming humanities scholarly practices to better locating and addressing research problems (M. Gold, 2012; M. Gold & Klein, 2019); while disputes accused those methods of disrupting the humanities identity as a discipline. Besides such debates, literature has also reported challenges in cross-field collaborations among researchers in digital humanities projects (Edmond, 2016; Flanders, 2012), making it increasingly important to investigate ways to achieve more in-depth, effective collaborations among DH communities, especially between the humanities and STEM researchers. As a result, non-verbal representations of knowledge are a central site to learn more about the existing tension and accompanied opportunities, as many representational forms above are inherently data-oriented and at the heart of DH and humanities-STEM collaborations (Drucker, 2020). It was only recent when the visuality in research started to receive scholarly attention from the digital humanities communities, calling for a deeper engagement with visual elements in both research processes and outcomes of digital humanities (Drucker, 2015, 2020; Münster & Terras, 2020).

In sciences, however, visuality has been a topic of extended examinations, where scholars frequently emphasize the concept of *inscriptions* and discuss their roles in shaping scientific fields. *Inscription* is commonly defined as "the material signs and artifacts of scientific production embodied in some medium" (Roth & McGinn, 1998). Naturally, this is a concept that covers a broad spectrum of non-verbal forms, including but not limited to all sort of information visualizations, tables, diagrams, and equations. Serving as an umbrella term for all non-verbal representations in scientific outputs and practices, inscriptions play significant roles in scientific research (Latour, 1990). Inscriptions function as "immutable mobiles," which can travel across research communities and at the same time retain stable meanings, making them a strong device in the construction and communication of scientific knowledge. Inscriptions also "enable scientists to interact with complex phenomena and convey important evidence not observable in other ways" (Latour, 1990). Besides their roles in knowledge production, an effective use of inscriptions also increases the persuasiveness of scientific literature (Shapin et al., 1985), which makes them a popular pedagogical tool in science education (Dimopoulos et al., 2003; Evagorou et al., 2015). All these studies explain the heavy uses of such inscriptions, especially data visualizations (or graphs) and tables in scientific research reported by empirical studies (Caissie et al., 2017).

Despite the importance of inscriptions in the scientific enterprise, there is a shockingly lack of empirical, quantitative efforts to investigate how inscriptions are used in DH research outputs, despite the scholarly efforts to advocate for a distinct, humanistic approach to visualizations (Champion, 2016; Drucker, 2011; Jessop, 2008). This gap is a missed opportunity to examine empirically how DH involves various research domains with distinct epistemologies and methodologies, especially because inscription creation and usage are deeply rooted in the data-driven paradigm of scientific research (Manovich, 2015). To fill this literature gap, this paper aims to provide a community-based, quantitative examination of inscriptional use in digital humanities publications and discuss the ways in which humanities and scientists working in the field approach inscriptions.

Specifically, we pursue the following research questions in this study:

**(1) What types of visual inscriptions are used in DH publications?**

The purpose of this question is to provide a typology and a comprehensive, descriptive survey of all the inscriptions in DH journal publications. We used manual coding to identify all inscriptions used in a selected sample of DH publications. Based on our sample, we modified and expanded an existing classification system of scientific inscriptions and applied the scheme on all inscription we found. Using the results of classification, we investigated how different types of inscriptions are used in the selected DH corpus. This survey represents the first piece of empirical evidence of how inscriptions are used in DH publications and the understanding of how the uses of inscriptions change over time.

**(2) How are these inscriptions used differently by researchers from various knowledge domains?**

This question aims to look at the different inscription practices and the potential tensions between research traditions in the DH field, especially the humanities and STEM, as the evolution of digital humanities has greatly benefited from the constant learning, critique, and incorporation from STEM research conventions and practices, such as information technologies, data science, or statistical approaches (Alvarado, 2012; Bradley et al., 2018; Bradley, 2019; Fitzpatrick, 2012). We manually classified all sampled articles into different knowledge domains, including *Humanities* and *STEM*, based on the affiliation information of their authors. Building upon the domain classification, we evaluated how inscriptions are approached by researchers from the humanities and STEM, under different collaboration situations, so as to understand the dynamics of inscription usage in the "battleground" of the DH field.

**(3) Which journals do different DH communities prefer to publish in?**

This question aims to examine the roles played by journals as a venue of the cross-disciplinary inscription-use patterns addressed in the previous research question. We selected four major journals that are the most heavily represented in our data sample and provided quantitative analyses on the inscription use and the composition of research domains. These journal-level patterns were then compared with those from our whole sample to understand the different roles played by these journals in the DH landscape.

This study makes important contributions to the digital humanities scholarship and studies on scientific non-verbal languages. First, it offers an empirical, comprehensive overview of the visuality in DH based on journal publications and our modified classification system. This greatly expands our existing knowledge on the disciplinarity of DH communities and will serve as an important baseline result for any further empirical research on similar topics. Second, by examining different visual strategies used by various DH communities, especially the humanities and STEM researchers, this study raises a novel research approach to analyzing non-verbal or visual representations in research outputs from the perspective of cross-disciplinary relationships, one that has profound implications to the fields of science studies.

## Literature Review

### Visual Representations in Science

Visualization is an important component in the scholarly communication system and has attracted great attention by the field of science and technology studies (STS), a sociologically oriented research field dedicated to investigating sciences. From the onset of STS as a research field, researchers have identified inscriptions produced during the research process as an important means to carry laboratory traces and play key roles in the production of scientific knowledge (Latour, 1987; Lynch & Woolgar, 1990).

The power of inscriptions lies on the fact that they are both inherent immutable and mobilizable, or, what Latour called the "immutable mobiles" (Latour, 1990). Being an immutable mobile means that an inscription possesses what Ivins (1973) called the trait of "optical consistency," where its internal properties cannot be easily modified after the inscription is produced (i.e., the immutability). Despite such consistency, the inscription can also embody the scientific reasoning process and practices, which enables the communication of laboratory observations and discoveries, persuades the general audiences and dissenters, and eventually mobilizes the traces of laboratory practices into widely accepted scientific knowledge. Such attributes give inscriptions unique rhetorical power that can bridge the mental, material and textual worlds and transforms laboratory traces into seeable evidences (Latour, 1990). Generating evidences that are visible is the holy grail during the whole research process from the practice-oriented view of sciences (Latour, 1987), because being able to see is the most basic foundation of scientific objectivity (Daston & Galison, 2007). A classic example of this argument is Shapin and Schaffer's (1985) famous recount of the air pump experiment, where Robert Boyle transformed the experience of witness into literary forms (or the *literary persuasion technique*). We can certainly find the reincarnation of this strategy in contemporary scientific artifacts and norms, such as including various types of visual inscriptions to scientific publications to make one's claims more appealing and credible (Richards, 2003; Shapin, 2007).

Despite this general view of how inscriptions contribute to scientific research, the relationship between inscriptions and specific research contexts is nevertheless contingent on many other factors, especially the level of data-drivenness of a research community (Cleveland, 1984; Smith et al., 2000), as inscriptions and especially visualizations are largely the outputs of data involved in scientific research. This leads to the inevitable conclusion that visualization is not perceived and used consistently across research disciplines (Gross & Harmon, 2014; Lynch, 1990; Myers, 1990). Moreover, as all visualization techniques are designed and used for specific data and task types (Shneiderman, 1996), we need a functional scheme to analyze how visualizations are plugged into research contexts to fulfill certain goals. For example, Lynch (1990) made the distinction between *selection* and *mathematization* as two important mechanisms in which the natural order is transformed or abstracted into scientific knowledge: *selection* refers to the simplification and schematization of objects of study while *mathematization* means using such visualization techniques to impose mathematical order above the natural objects. Such transformations, particularly the order from *selection* to *mathematization*, are critical to the production of scientific knowledge, as scientific knowledge is the "universal" pattern that is beyond the natural world saturated with uncertainties, details, and local contexts (Knorr-Cetina, 1981; Myers, 1990).

Given the strong epistemic roles played by inscriptions in knowledge production and communication, they have been used as a quantitative indicator to evaluate the level of scientificity of research fields. Among these studies, frequency per page (i.e., the count of inscriptions used per page, Butler, 1993) and fractional graph area (i.e., the ratio of publication area devoted to visual inscriptions; Cleveland, 1984; Smith et al., 2000) are two indicators developed in earlier works. These studies have found a general positive correlation between the amount of graphs used in publications and the extent to which a research field is regarded as scientific (Cleveland, 1984; Smith et al., 2000), or the famous "graphism thesis" proposed by Latour (1990). Moreover, Arsenault et al. (2006) went further to demonstrate that non-visual inscriptions, such as tables or equations, do not have the same effects on the hierarchy of sciences with graphs; instead, they are inversely correlated with the "hardness" of scientific fields. All these studies represent a tradition of scientometrics that focuses on scientific graphs, to which the present study aims to contribute. But more importantly, we conducted this research in the site of digital humanities to understand how humanistic research approaches and data-driven techniques are combined and combated in this unique cross-disciplinary field.

**Humanistic Visualization**

The use of visual information as instruments in humanities research is not a new phenomenon: the large body of scholarship in cartography, for example, has always been closely related to humanities disciplines, such as the history (Jessop, 2006). However, computer-assisted visualizations in digital humanities only started to be widely received among DH communities in the 2000s, when new visualization techniques were developed and applied to facilitate the *distant reading* of texts (Moretti, 2005; Sinclair, 2003).

With such a rise of visualization in digital humanities, scholars have developed distinct thoughts for more ideal uses of the visual language. The idea of *humanistic visualization* has emerged as an attempt to establish a new visual language paradigm in humanities, as contrast to *scientific visualization*. Jessop (2008) first emphasized that a humanistic visualization should embody the interpretation process of humanities research and not be used simply as a form of data display towards the end of the research. This idea has been echoed by several scholars in their works, such as the non-representational approach to *humanities visualization* (Drucker, 2018) and the idea to model humanities visualization as a sandcastling process (Hinrichs et al., 2019). Manovich (2011) proposed the concept of *direct visualization* that aims to create visual representations based on the actual visual media objects or their parts. Opposite to the *abstraction principles* of scientific visualization that aim to offer clear and accessible messages from the data and the natural phenomena, direct visualization values and preserves the complexity and the original forms of the humanities data. Furthermore, Drucker (2011) raised the concept of *capta* to acknowledge the "situated, partial, and constitutive character" of humanities inquiries and knowledge production. Compared with the widely used concept of data, capta is a *given* rather than a *taken*, which captures the interpretative nature of humanities research approach (Drucker, 2011). Extending from the conceptual discussions, scholars have also identified *uncertainty, ambiguity,* and *subjectivity* as crucial issues to consider when applying visual techniques to model interpretations in humanities research (Bradley et al., 2018; Champion, 2016; Drucker, 2020), ostensibly contrasted to how data visualization is perceived in sciences.

These are, however, merely theoretical discussions that do not reflect the inscription practices among DH research communities; and we need empirical evidence to develop an understanding of the *status quo* in DH visualization and to further support the design of humanistic visualizations. With the recent increase in the scale of specialized workshops, conferences, and publications that focus on visualization in DH (Bradley et al., 2018;  Jänicke et al., 2017), we have an unprecedented opportunity to grasp DH visualization in an empirical manner. And this research opportunity has started to attract scholarly attentions. Jänicke et al.'s (2017) state-of-the-art report on textual data visualization techniques demonstrated an early attempt to provide an overview of the DH visualization with empirical analyses. Their 5-year extensive work analyzing papers published in both representative DH and visualization journals classified data visualization techniques based on their supports for either close or distant reading of texts. Weingart & Eichmann-Kalwara (2017) further went beyond the textual data visualization and empirically demonstrated the increasing scholarly attention to visualization by analyzing abstracts published at the DH conferences from 2004 to 2015. Beyond these studies, no empirical work has provided detailed accounts of the usage pattern of DH visualizations, especially the different approaches to visualization among humanities and STEM scholars and how the differences impact collaboration of the two research communities on DH work.

## Methods

### Data

In this project, we used DH journal publications to pursue our research questions. An identified challenge in the quantitative examination of DH publications is the fact that humanities journals are poorly indexed in major bibliographic databases, such as the Web of Science and Scopus (Norris & Oppenheim, 2007). However, Spinaci et al.'s research work (2020) identified a core exclusively DH journal list from the Crossref database, which makes the present research possible. In our study, we retrieved all English-language research articles in the above list that (1) have a valid title and DOI and (2) are published between 2011 and 2020. We only selected articles with a valid title and DOI so that we can acquire full texts of these articles later. We selected this specific publication year window because (1) ten years are a long enough period to reveal a meaningful temporal trend from the analyzed materials and (2) this year window covers major developments of DH visualization, such as new theoretical contributions, new visualization tools, and new conferences and workshops focused on DH visualization (Bradley et al., 2018; Moretti, 2005; Sinclair, 2003). 1/15/21 9:33:00 AM

A total of 1,566 journal articles were acquired on July 30, 2020. From all retrieved articles, we randomly selected 300 articles as the sample of the present study and manually downloaded the PDF files of their full texts from journals websites.

It should be noted that while all retrieved articles were marked as English-language research articles by Crossref, we found that some of them are written in a different language or is not a research article. We removed these articles during manual coding, given the difficulties in validating this information based on metadata alone.

**Identification and Classification of Inscriptions**

For all selected articles, we manually identified what inscriptions are used in the publication and classified all inscriptions into a category based on the visual forms they take. Two coders independently reviewed the downloaded PDF files of selected articles. To classify inscriptions, we developed the classification system of scientific inscriptions proposed by Arsenault et al. (2006), based on the characteristics of DH scholarship and the sample we examined. Our final classification scheme is listed in Table 1, with working definitions for each category.

**Table 1: Classification scheme and definitions of inscription types**

| Category 1 | Category 2 | Definition |
|---|---|---|
| Graphs | Graphs | Graphic representations of empirical and quantitative data (Arsenault et al., 2006; Azzam et al., 2013) |
| Non-graph illustrations | Collages | Combination of multiple types of visual forms into one single display (Montgomery, 2002) |
| | Diagrams | Spatial arrangement of elements to convey information and show hierarchical organizations (Arsenault et al., 2006; Butler, 1993) |
| | Illustrations | Pictorial representation of natural objects and phenomenon (Arsenault et al., 2006) |
| | Maps | Visual representation of geographical location information (Arsenault et al., 2006; Grant, 2019) |
| | Montage | Combination of related visual artifacts together to create a larger work in its entirety (Manovich, 2012; Manovich, 2012) |
| | Photographic images | Isomorphic, realist representations of natural objects and phenomena (e.g., photos, computer screenshots) (Arsenault et al., 2006; Myers, 1990) |
| | Simulations | Virtual reproduction of a physical scene (e.g., an archaeological site) or virtual blueprints for an imagined artifact created with computer programs (e.g., virtual reality images) |
| Non-visual illustrations | Codes | Textual display of digitalization or computational procedures (e.g., algorithms, programming codes) |
| | Equations | Mathematical expressions that were |

| | | set off from the body of the text in the style of a block quotation (Arsenault et al., 2006) |
|---|---|---|
| | Tables | Arrays of information consisting of rows and columns and set off from the body of the text (Arsenault et al., 2006) |
| | Texts | Encoded representation of textual information (e.g., HTML, XML) |

We retained most of the categories in Arsenault et al.'s (2006) original scheme, including the most frequently used inscriptions based on previous studies, tables and graphs (Caissie et al., 2017), as they also apply to our paper sample and more generally fit the humanistic approaches to inscriptions. Moreover, we also retained the top-level categories, i.e., *graphs*, *non-graph illustrations*, and *non-visual illustrations*, as a general organizing framework even though they are not used in our analysis. Like how the category is treated in classic literature, *graphs* refer to the broad family of information visualization as graphic representations of empirical data (Azzam et al., 2013), comparing to some other similar visual displays in the category of non-graph illustrations. This definition is especially helpful in our distinctions between *graphs* on the one hand, and *maps* and *diagrams* on the other hand, as both of which share some visual components that are central to their functions.

Moreover, we added a few categories to the classification scheme to better accommodate the domain of digital humanities and our paper sample. **Our changes are along two directions: (1) inscription types that are emerging and recent and (2) those that are specific to DH research.**

In the first category, we added *simulations* (under NGI) and *codes* (under NVIs). With the rise of new visual technologies and techniques such as the virtual and augmented realities, simulations are frequently applied in papers on virtual cultural heritage and archaeology to showcase the design of a virtual program or the re-modelling of a physical site (Pujol-Tost, 2017). Similarly, as the demonstration of computational procedures of a task (e.g., algorithms or blocks of programming codes; Hsiang et al., 2012), a frequent use of codes suggests the increasing emphasis on computational methods in DH.

The second category of our modification focuses on raising categories specific to DH research, such as the expansion on the photographic images to include *screenshots*. As we observed from our data sample, screenshots are frequently used in DH publications to highlight essential features and functions of a digital tool or interface (Bartalesi et al., 2018). This trend demonstrates the emphasis on digital infrastructure development in DH research, a research area that has also been widely discussed in DH scholarship (Schreibman et al., 2016). Moreover, we also included the *texts* category (e.g., XML, HTML) under NVIs, whose development is associated with the enduring need in DH research to work with textual information in digital environment. For this category, we only counted those textual representations in dedicated chunks in the publication, to distinguish them from regular texts in the paper.

Our manual coding focuses on the second-level categories in Table 1. The inter-coder agreement between the two coders is 0.887, which reaches the level of "very good" based on Landis and Koch's classic recommendation (1977). All differences between the coders were resolved before analysis.

## Measurements of Inscription Use

Earlier works have adopted frequency per page (i.e., the count of inscriptions per page; Butler, 1993) and fractional graph area (i.e., the ratio of publication area devoted to visual inscriptions; Cleveland, 1984; Smith et al., 2000) as quantitative indicators of how graphs are used on the paper level. We argue that the second indicator is not suitable for publications that are born digital, so we decided to use a measurement that is similar with frequency per page, *total count of inscriptions*. We did not standardize the frequency of inscriptions by page numbers of articles because some papers were published as web pages, whose page numbers are essentially impossible to be counted. Moreover, we also used the *complexity of inscriptions* as a supplementary measurement. Both measurements are explained below:

- **Total count of inscriptions**: This parameter measures the total count of inscriptions used in a publication. Similar with previous studies, we hypothesize that the more frequently inscriptions are used, the more connected to sciences a publication (as well as its authors and journal) is.

- **Complexity of inscriptions**: This parameter measures how many types of inscriptions are used in a publication. We again hypothesize that the more diverse inscriptions are used in a publication, the more strongly connected to sciences the publication as well as its authors and journal tend to be.

## Identification of Author Affiliation and Domain

We developed a classification scheme to assign all authors into research domains that are relevant to our research question, i.e., how researchers from the STEM and humanities fields (among others) are approaching inscriptions in DH journal publications. Our classification is shown in Table 2. We specifically distinguished authors working in an academic institution from those who are not. The latter category, named *non-academic*, includes libraries, archives, schools, and companies that are related to DH research. For academic domains, we classified authors into *STEM*, *Social sciences*, and *Humanities*. Over the past few years, there has been a growing trend to establish DH-focused institutions in academic settings, such as the Department of Digital Humanities at King's College London and the Institute for Advanced Technology in the Humanities at the University of Virginia (Fraistat, 2012). For researchers who work at these DH-focused departments or institutions, we used *Interdisciplinary institution* instead of one of the aforementioned domains. Our classification scheme is very similar with the one adopted by Sula and Hill (2019), although we grouped specific disciplines into domains (e.g., the *humanities* and *STEM*) based on our analytical purposes and a broader scope.

**Table 2: Classification scheme for author domains**

| Author Domain |
| --- |
| Humanities |
| STEM |
| Interdisciplinary institutions |

| Social sciences |
| --- |
| Non-academic |

One coder independently reviewed the information of all authors in our sample and applied the classification scheme. We relied on the following sequence of information sources to determine an author's affiliation: (1) information supplied in the article, (2) the author's personal websites or institutional pages, and (3) other information sources with such information, such as ResearchGate and Academia.edu. The information in the article is prioritized because it supplies the timeliest information. However, we still used other sources because a large number of articles do not have any author affiliation information and we made the assumption that authors' research domains do not change very easily. Any issue in the coding was discussed between the two authors and solved before analysis.

After the coding, there are a number of authors whose affiliation information cannot be found. Given the centrality of this information to our study, those papers that do not contain any author with valid domain information were removed from our sample, which leads to the removal of one paper from our sample. More details about our final sample are discussed in the next subsection.

**Descriptive Analysis**

Our final sample includes 252 articles, after the removal of papers that (1) are not written in English, (2) are not research articles, and (3) do not have any author with any valid domain information. Below is the summary of all papers by their publication years. Figure 1 shows the distribution of publication years of all these papers.

**Figure 1: Count of articles by year**

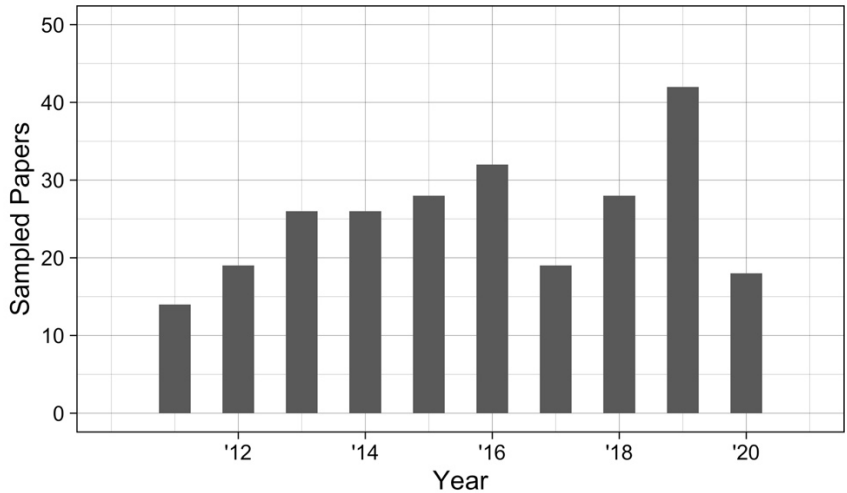

Table 3 shows the total counts of all inscription types identified in our sample. Tables, graphs, and photographic inscriptions are the three most frequently used types. On the other side of the spectrum, montages, illustrations, and maps are among the least frequently used categories based on our classification scheme.

**Table 3: Count of inscription types**

| Type | Count |
|------|-------|
| table | 678 |
| graph | 483 |
| photographic inscriptions | 350 |
| diagram | 281 |
| equation | 237 |
| text | 237 |
| simulation | 71 |
| collage | 49 |
| code | 44 |
| map | 29 |
| illustration | 11 |
| montage | 1 |

Table 4 summarizes the number of articles published in different journals. Four journals contain more than 20 articles in our final sample, and they are used as the primary sources for journal-based analysis to be presented later.

**Table 4: Count of journals**

| Journal | Count |
|---------|-------|
| Digital Scholarship in the Humanities | 82 |
| International Journal of Humanities and Arts Computing | 37 |
| Journal on Computing and Cultural Heritage | 33 |
| Literary and Linguistic Computing | 29 |
| Digital Studies/Le champ numérique | 19 |
| Journal of the Text Encoding Initiative | 18 |
| Frontiers in Digital Humanities | 17 |
| Digital Medievalist | 7 |
| Journal of the Japanese Association for Digital Humanities | 4 |
| International Journal of Digital Humanities | 2 |
| Journal of Cultural Analytics | 2 |
| Revista de Humanidades Digitales | 2 |

In total, we identified 611 author instances with valid domain classification. We further classified all papers into our defined domains based on the first author. Table 5 presents a summary of the classification results, including the number of first-author papers and the total count of author instances. From the table, humanities and STEM authors are the major contributors in DH journals. It is also important to note that researchers in innately DH institutes are also heavily represented in DH publications and they may be underrepresented in our sample because of the fact that newer DH institutions are still being created. Moreover, even though humanities researchers contributed more first-author papers in our sample, there are more instances of STEM authors, partly because of the longer authors lists in STEM-first-author papers.

**Table 5: Summary of author domains**

| Domain | First-author papers | Single-authored papers | Total authors | Mean number of Authors in first-author papers |
|---|---|---|---|---|
| Humanities | 104 | 53 | 208 | 1.91 |
| STEM | 82 | 19 | 248 | 3.24 |
| Interdisciplinary institute | 32 | 14 | 75 | 2.22 |
| Social Sciences | 20 | 10 | 38 | 2.1 |
| Non-academic | 14 | 5 | 42 | 2.36 |

# Results

## How are Inscriptions Used Over Time?

To understand the distribution of inscriptions across our sample, we calculated the mean number of inscriptions per paper and inscription complexity over time, as shown in Figure 2. The graph illustrates a slight increase in both variables during the past 10 years, despite the yearly fluctuations. For the number of inscriptions per paper, the number has increased from 8.5 to 11 from 2014 to 2020, whereas the mean inscription complexity has remained around 3 since 2015.

**Figure 2: Inscriptions per paper (top panel) and inscription complexity (bottom panel) over time in the sample**

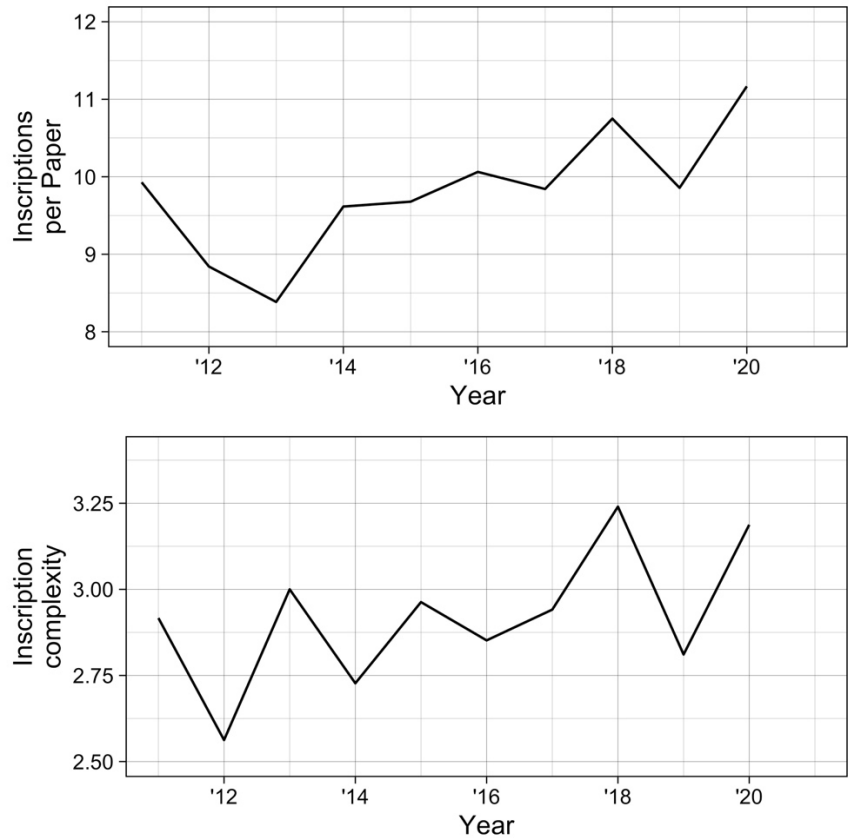

In Figure 3, the temporal trend of all inscriptions is broken down into each category, with collages, codes, maps, illustrations, and montages combined into a single *others* category, given their low frequencies. The y-axis of the graph shows the mean recurring frequency of each inscription type on the paper level. Some inscription types show rather radical year-to-year changes, which is attributed to the relatively small occurrences of these types in the sample. However, two patterns emerge from the results. First, equations are increasingly used in our sample during the publication window while *text* shows a strong opposite trend. This supports the general idea that DH research is becoming more mathematized and less textualized. Second, the use of some other inscription types, especially graphs and diagrams, has also slightly increased since 2012. We calculated coefficients of the publication year to the outcome variable based on Figure 3. The slope values for graphs and diagrams are 0.064 and 0.033, respectively, comparing to 0.162 for equations. Besides these three inscription types, all other types, except for *text*, have only been mildly increasingly used over time, with the coefficients value lower than 0.01. Against this general rising trend, *text* is the only category with a negative slope value, which is -0.152.

**Figure 3: Mean numbers of inscription types over time**

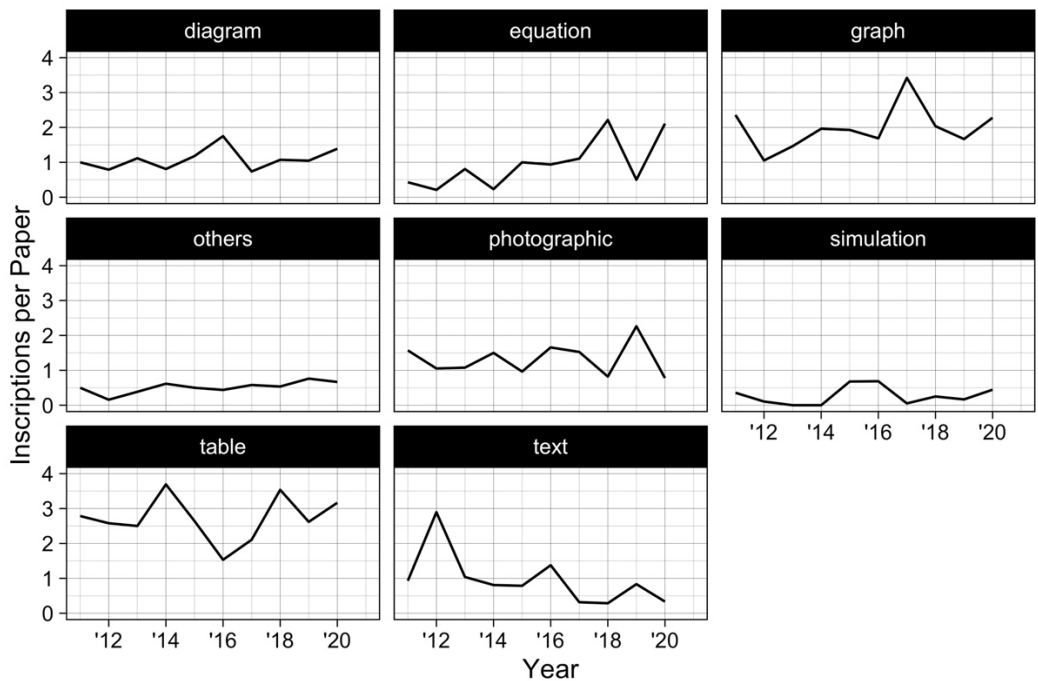

## How Are Researchers from Different Domains Using Inscriptions?

A central interest of this work is to analyze how researchers from different domains, especially STEM and humanities, use inscriptions in DH journal publications. Table 6 shows that papers with STEM researchers as the first author (*STEM-first-author papers*) are more inscription-intensive in terms of both count and complexity, while all other categories except *non-academic* have similar statistics. One of the reasons for the really low number of inscriptions used in the *non-academic* category may be its low number of papers.

**Table 6: Inscription Use among First-Author Domains**

| First-author Domain | Inscriptions per paper | Inscription complexity |
|---|---|---|
| STEM | 13.16 | 3.34 |
| Interdisciplinary institute | 9.44 | 2.59 |
| Social Science | 8.65 | 2.6 |
| Humanities | 8.32 | 2.06 |
| Non-academic | 3.71 | 1.14 |

Figure 4 shows the number of each major category of inscriptions in these different domain-oriented paper groups. From this graph, we found that nearly all inscription types are the most heavily used in STEM-first-author papers, except for *text*. However, while graphs and tables are the most intensively used in STEM papers, there is normally not a big difference between STEM and other groups.

**Figure 4: Inscription Types across First-Author Domains**

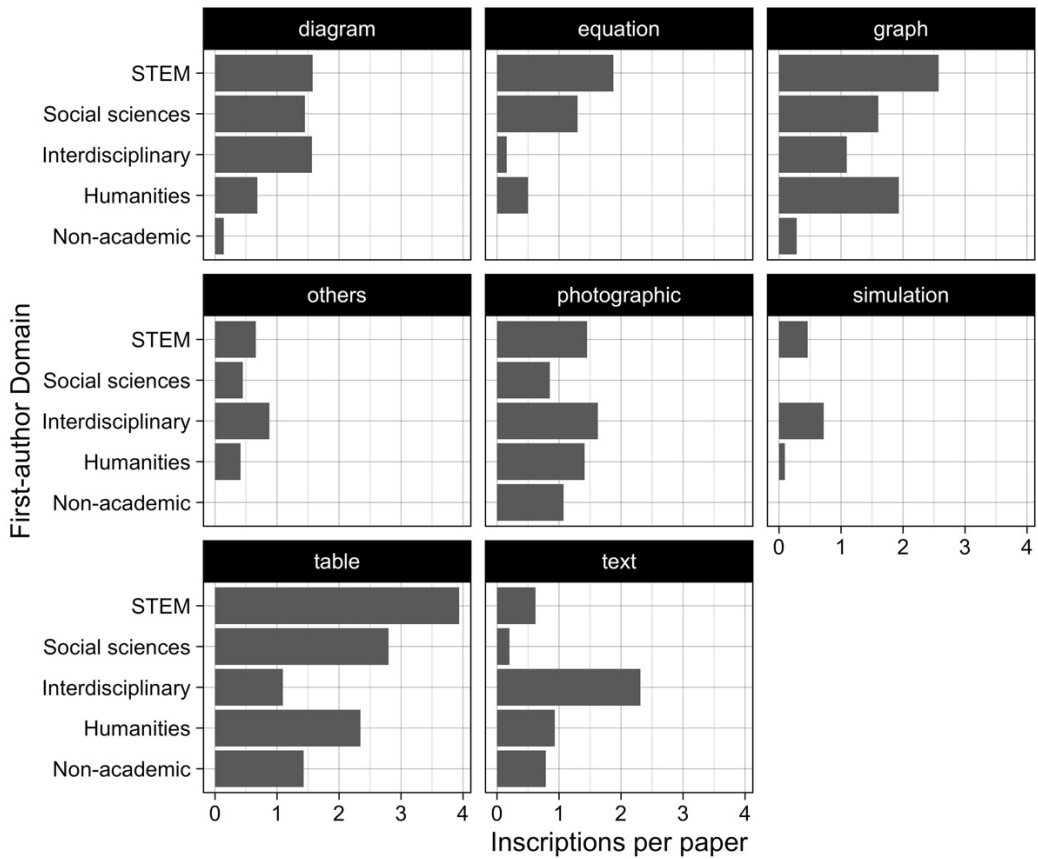

One potential reason to explain the not so obvious differences in the inscription usage between STEM and other domain groups is that the first author alone is not the sole representation of the domain attribute of a paper: only 40% of all sample papers are single-author papers and this ratio is only 52.5% in the Humanities group. To understand how collaboration between authors from different domains contribute to the paper's domain attribute, we examined inscription usage in publications collaborated by STEM and humanities authors. Based on (1) the first author of the paper (STEM or Humanities) and (2) whether the paper is collaborated by STEM and Humanities authors, we classified all papers with authors whose first author is from either humanities or STEM into the four categories shown in Table 7. While it is obvious that most publications are not collaborated between these two groups of researchers, there is a much higher ratio of collaborative publications among STEM-first-author papers than the humanities group.

**Table 7: Collaboration between Humanities and STEM Authors**

| First-author Domain | Collaboration | Non-collaboration |
|---|---|---|
| STEM | 16 | 66 |
| Humanities | 12 | 92 |

Despite the small number of publications collaborated by STEM and humanities researchers, we compared these publications in terms of the two inscription-use measurements adopted in this study. From Table 8, we found that no matter which domain the first author comes from, when the paper is collaborated between authors from the humanities and STEM fields, the inscription usage tends to be in the middle of the two boundaries. However, *STEM-collab papers* are still more inscription-intensive than *humanities-collab papers* on both measurements.

**Table 8: Does Cross-Field Collaboration Affect Inscription Use among Humanities and STEM Domains?**

| Domain | Collaboration with each other? | Inscriptions | Inscription complexity |
|--------|-------------------------------|--------------|------------------------|
| STEM | Non-Collab | 13.29 | 3.39 |
| STEM | Collab | 12.62 | 3.12 |
| Humanities | Collab | 10.75 | 3.17 |
| Humanities | Non-Collab | 8 | 1.91 |

**What Roles Are Journals Playing in the Inscription-Domain Relationship?**

In light of the relationship between researchers' domains and inscription use in DH publications presented above, in this subsection, we aim to understand the roles of journals in this landscape. Table 9 shows the summary of mean number of inscriptions and mean inscription complexity on the paper level for the four major journals with at least 20 publications contained in our sample. The results, again, show a quite strong correlation between these two measurements of inscription use in DH publications. But more importantly, the *Journal on Computing and Cultural Heritage* stands out as the most inscription-intensive journal in our list of journals.

**Table 9: Use of Inscriptions across Journals**

| Journal | Inscriptions | Inscription Complexity |
|---------|--------------|------------------------|
| Journal on Computing and Cultural Heritage (JCCH) | 16.7 | 4.42 |
| Digital Scholarship in the Humanities (DSH) | 10.52 | 2.51 |
| Literary and Linguistic Computing (LLC) | 8.97 | 2.07 |
| International Journal of Humanities and Arts Computing (IJHAC) | 6.87 | 2.02 |
| Others | 7.65 | 2.14 |

We also evaluated how different inscription types are distributed across these journals. Figure 5 shows the composition of inscriptions in the four journals as well as all the other journals. On one hand, we found that the *Journal on Computing and Cultural Heritage* has more photographic inscriptions, equations, diagrams, and simulations, many of which are indicative of an algorithmic or mathematic style of research. On the other hand, consistent with our previous results, the uses of tables and graphs are quite similar across journals.

**Figure 5: Inscription Types across Journals**

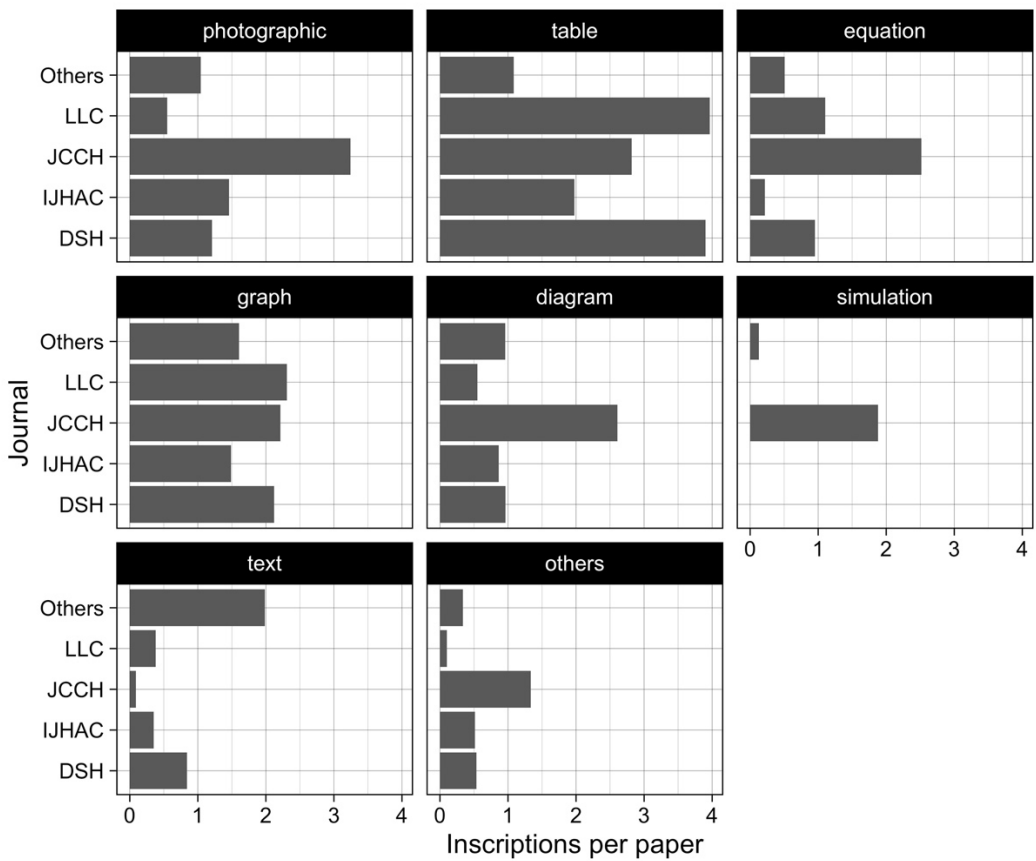

Such differences in the use of inscriptions can also be explained by the composition of researchers from different domains in these journals. Table 10 summarizes the ratios of papers in each journal that have the first author from humanities, STEM, and other domains. The table shows that The *Journal on Computing and Cultural Heritage* has a significantly higher ratio of articles with STEM first authors than other journals. We can see a strong correlation between the composition of authors in each journal and its inscription use, shown in Table 8. The *Journal on Computing and Cultural Heritage* clearly has the highest ratio of papers with STEM first authors and a much heavier use of inscriptions than other journals. One exception in this general relationship is the *Digital Scholarship in the Humanities*: even though it has one of the lowest STEM first-author paper ratio, its inscription use is more intensive than the other two journals.

**Table 10: Paper and Author Distribution across Journals**

| Journal | Ratio of Humanities papers | Ratio of STEM papers | Ratio of other papers |
|---|---|---|---|
|  |  |  |  |
| DSH | 48.8% | 30.5% | 20.7% |
| IJHAC | 40.5% | 29.5% | 29.7% |
| JCCH | 15.2% | 69.7% | 15.2% |
| LLC | 34.5% | 44.8% | 20.7% |

## Discussion

### Inscriptions Used by DH Publications

In this study, we revised a well-received classification scheme of scientific inscriptions proposed by Arsenault et al. (2006) to include inscription categories meaningful to DH research. Focusing on inscription categories that are specific to DH (such as *photographic images* and *texts*) and those that are emerging (such as *simulations* and *codes*), we were able to construct a DH-focused inscription classification scheme that can well represent inscriptions in the latest DH publications and has the potential to be reused as a research instrument in future studies.

Based on our revised classification scheme, we investigated how all inscriptions and individual categories from our sample are used over time. We find a general increasing trend in terms of both the total frequency of inscriptions and inscription complexity in DH publications among all inscriptions. We believe this is a strong evidence of the growing importance of visuality in DH scholarship over the past decade. And based on Latour's graphism thesis (Latour, 1990), it seems that the DH field, in general, is becoming more "scientific" from the early 2010s.

This general conclusion is further supported by our findings regarding individual inscription categories. We found that all inscription categories, except for text, are increasingly used in our paper sample. Equation is the most increasingly used inscription type over the past decade, followed by diagrams and graphs. We can argue that all of these three inscription categories are indicative of computational and mathematical research methods, as compared to text, the only category that has been clearly decreasingly used. All these facts suggest that the DH as a general research field has been increasingly reliant on the scientific paradigm of research, comparing to more traditional textual representations of knowledge, a trend that is contradictory to the argument that "digital humanities is text heavy, visualization light, and simulation poor" (Champion, 2016).

### How Are Inscriptions Used Differently by Humanities vs. STEM Researchers?

Visualization in DH has proved to be a research area with engagement from both humanities and visualization scholars (Benito-Santos, 2020; Stefan Jänicke, 2016). Collaborations among the two research communities are inevitable. However, no research has explored the impacts of such collaborations on the research practices in DH, especially those related to inscription use. Taking the metaphor of "DH as a battleground," our research investigated how inscriptions are used differently by researchers from distinct domains.

Our results show that STEM researchers use the greatest number of inscriptions per paper and demonstrated the highest degree of inscription complexity when they are the first author of the papers, as comparing with researchers from other domains. In comparison, papers with the first author from humanities, social sciences, and DH-focused institutions all have very similar statistics in terms of inscription use. This supports the thesis that inscription usage is positively connected to how scientific a field is, as validated in previous empirical studies (Cleveland, 1984; Smith et al., 2000). In terms of the inscription types, we find a general correlation between the first author domain and the inscription types that are more strongly connected to sciences from the previous research question. For example, STEM researchers have the highest uses of equations, diagrams, and graphs among researchers from all domains, whereas *text* is more frequently used by researchers from humanities and non-academic institutions. We should point out that not all results on the level of individual inscription type are consistent with our expectations, which may be largely due to the fact that the sample size on the inscriptions and domain levels are quite small. This limitation of our research warrants a future study using a larger sample size.

Moreover, the usage pattern of inscriptions is influenced by the collaboration among all authors from various domains. While both the humanities and the STEM researchers tend to use inscriptions distinctly when working alone, collaboration between them reconciles this difference. This is an interesting finding that sheds light on the cross-disciplinary collaboration dynamics in DH that has not been investigated by more theoretical-driven research. We can draw a preliminary conclusion from this finding, which needs future validations: the very different research approaches used by STEM and humanities researchers are quite evenly blended in such STEM-humanities collaborations.

## DH Journals as Distinct Epistemic Communities

Results from the journal analysis demonstrate a clear sense of community in DH in terms of the inscription usage. Different research communities, especially the STEM and humanities researchers, prefer different forms of visual communications and publication venues; and such preferences have shaped distinct epistemic cultures around journals. *Journal on Computing and Cultural Heritage (JCCH)* is a journal in which STEM researchers prefer to publish and one that embraces highly collaborative DH works. In terms of the inscriptions, *JCCH* demonstrates the greatest number of inscriptions on average and the highest level of inscription complexity among all the journals being examined. The use of photographic images, diagrams, simulations, and equations exceeds all the other journals; but *text*, by contrast, is not a preferable inscription type in *JCCH*. Such distinct use of inscriptions and domain composition in *JCCH* can be explained by the epistemic positioning of this journal. Launched in 2008, *JCCH* publishes work on the "use of information and communication technologies (ICTs) in support of Cultural Heritage." Compared with the other three journals, *JCCH* has a particular focus on digital cultural heritage and a stronger technological emphasis.

Humanities researchers, in comparison, prefer to publish in more comprehensive DH journals that share the humanities tradition, such as the *Digital Scholarship in the Humanities (DSH)* and the *International Journal of Humanities and Arts Computing (IJHAC)*, which originated from a humanities discipline and was developed them to cover all aspects of computing and information technology applied to arts and humanities research (Edinburgh University Press, n.d.; Oxford

Academic, n.d.). They also tend to use fewer and less complex inscriptions – mostly the tables and graphs – and appear to have the preference to work with smaller teams (in this case, the average author number for each published paper in these two journals remains around 2). Our journal analysis empirically illustrates the "battleground" condition in the DH field and reflect two major communities of practice between STEM and humanities researchers with regard to their use preferences for inscriptions.

## Conclusion

In the present paper, we offered a quantitative examination of how inscriptions, such as information visualizations and tables, are used in a selected sample of DH journal publications. Specifically, we investigated how the use of such non-verbal representations is influenced by the participation of researchers (as well as their collaborations and competitions, based on the battleground metaphor) from different domain in the DH field, all of whom have their own distinct research approaches and methods that are deeply connected to the inscription use. We manually coded and classified all inscriptions used in 252 randomly selected English-language research articles published from 2011 to 2020 in a list of core and exclusively DH journals as well as the domain information of all authors in these publications. Using such information, we examined the evolution of inscription usage in these DH journals and how the use of inscription is related to the domain from which authors come from. We found that inscriptions have been slightly more frequently used in DH publications over the past decade, which is supported by the increasing uses of nearly every type of inscriptions per our classification system, except for *text*. Moreover, equations, graphs, and diagrams, three types of inscriptions that are strongly connected to *mathematization*, are found to be more increasingly used in our sample than other inscriptions. Finally, we also identified a strong connection between inscription use and the author domains. A paper with STEM authors as the first author tends to have more inscriptions used than others and collaborations between STEM and humanities authors have similar inscription use statistics no matter who is the first author. The pattern between the domain of authors and inscription use is also shown on the journal level: The *Journal on Computing and Cultural Heritage* stands out as a unique publication venue in terms of the use of inscriptions and domain composition from other important DH journals, which points to the existence of different inscription-oriented sub-communities within the field of DH.

Our paper makes the following contributions to digital humanities and the quantitative science studies focusing on visuality. First, our paper is the first comprehensive empirical analysis on how visualizations or inscriptions are used in DH, based on our best knowledge. This bridges an important gap between the increasing interests in the topics of data visualization and the data-driven research approach among DH communities (Jänicke et al., 2017) and a lack of empirical research on such topics. Our work largely confirms that DH as a research field is becoming more computational, even though this change is gradual and slow. Second, our paper expands existing quantitative efforts on research inscriptions that are largely only focus on sciences. By drawing attention to the field of DH, a field that is highly cross-disciplinary and without clear methodological and theoretical boundaries with an array of divergent research traditions, our study revised the scheme of research inscriptions and identified interesting relationships between the use of non-verbal representations and collaborations between different research domains.

Despite its important and timely contributions, we acknowledge that the present paper has a few limitations, which should be addressed in future works. First, due to the labor-intensive coding process, this study is based on a relatively small sample, which prevented us from drawing more granular conclusions concerning how particular inscription types are used in various contexts, especially in collaborations between authors from other domains. Second, this work only focuses on one research field, the DH, which confines the depth of our conclusions. We are planning to undertake the next steps of this research project, which will be based on larger samples from multiple research domains, to gain deeper appreciation of the roles played by visuality in scholarly outputs and activities.